\documentclass[prl,aps,twocolumn]{revtex4}
\usepackage{graphicx}
\date{\today}
\begin{document}
\title{Solution of an infection model near threshold}
\author{David A. Kessler}
\email{kessler@dave.ph.biu.ac.il}
\affiliation{Department of Physics, Bar-Ilan University, Ramat-Gan, IL52900 Israel}
\author{Nadav M. Shnerb}
\email{shnerbn@mail.biu.ac.il}
\affiliation{Department of Physics, Bar-Ilan University, Ramat-Gan, IL52900 Israel}
\begin{abstract}
We study the Susceptible-Infected-Recovered model of epidemics in the vicinity of the
threshold infectivity.  We derive the distribution of total outbreak size in the limit of large population size $N$.  This is accomplished by mapping the problem to the first passage time of a random walker subject to a drift that increases linearly with time.  We recover the scaling results of Ben-Naim and Krapivsky that the effective maximal size of the outbreak scales as $N^{2/3}$, with the average scaling as $N^{1/3}$, with an explicit form for the scaling function.
\end{abstract}
\maketitle
Recently, Ben-Naim and Krapivsky\cite{BnK} (BN-K) studied the statistics of the size of an epidemic in the
Susceptible-Infected-Recovered (SIR) model\cite{Bailey,AM,Hethcote} when the infectivity is near its threshold value.
When the infectivity is below threshold, an outbreak quickly dies out, infecting some
finite number of individuals, essentially independent of the population size. Above the
threshold, the total average number of affected individuals reaches a finite fraction of the population. BN-K found that at threshold, the total average number of affected individuals is proportional to $N^{1/3}$,  and that there are essentially no outbreaks which infect more than of order $N^{2/3}$ victims.  While presenting an argument justifying these scaling laws, no analytic calculations for the distribution of outbreak sizes was given.  In this paper, we present an exact formula for this distribution, in the limit of large population sizes, which of course exhibits the scaling properties found by BN-K.  This calculation involves solving an auxiliary problem, namely the first-passage time statistics\cite{Redner} for a random walker released at $x=1$ to be absorbed at the origin, given a small leftward drift which increases in magnitude linearly in time.  This problem is one of the few such first-passage problems with time-dependent forcing\cite{Lindner} for which an analytic solution is available, and so is of independent interest.

We begin with a description of the SIR model.  The $N$ individuals in the population are divided into three subclasses: the susceptible pool, of size $S$; the infected (and infectious) class, of size $I$; and those recovered (and no longer infectible), of size $R$, with $N=S+I+R$. The disease is transmitted from an infected individual to a susceptible one with rate $\alpha/N$, so that
\begin{equation}
(S,I,R)\stackrel{\alpha S I/N}{\to} (S-1,I+1,R) .
\end{equation}
Infected individuals recover with a rate which we can take to be unity:
\begin{equation}
(S,I,R)\stackrel{I}{\to} (S,I-1,R+1).
\end{equation}
Of primary interest is the case where initially $S=N-1$, $I=1$, $R=0$, so that the outbreak is sparked by a single infected individual.  The outbreak of course terminates when the last infected individual recovers, and $I$ returns to 0.

This stochastic process is traditionally approximated (for large populations) by the classic SIR rate equations
\begin{eqnarray}
\dot{S}&=& - \frac{\alpha}{N} S I \nonumber\\
\dot{I}&=& \frac{\alpha}{N}SI - I \nonumber\\
\dot{R}&=& I
\end{eqnarray}
Since $S$ decreases monotonically, these equations are easiest dealt with by eliminating the time and considering $dI(S)/dS$, which is obtained by dividing the second rate equation by the first:
\begin{equation}
\frac{dI}{dS} = -1 + \frac{N}{\alpha S}
\end{equation}
with the solution
\begin{equation}
I = N-S + \frac{N}{\alpha}\ln (S/(N-1))
\end{equation}
It is clear that if $\alpha < N/(N-1) \approx 1$, the rate equation predicts that the number of infected individuals decreases monotonically in time (decreasing $S$), whereas if $\alpha$ is greater than this threshold, the number of infected individuals first rises, and as $S$ decreases, eventually
$N/S$ falls below $\alpha$ and $I$ falls till it hits 0 at $S_f$ satisfying $N-S_f \approx (N/\alpha)\ln(N/S_f)$.  Thus at the classical level, $\alpha=1$ marks the threshold between an infection that infects a finite percentage of the population and those that fail to spread.

To study the stochastic process, we adopt a similar strategy and eliminate time, focussing solely on the transitions between states. We characterize the system by the
number of transitions the system has undergone. In each transition the number of infected individuals either rises or falls by one, so that $I$ undergoes a kind of random walk.  After $T$ transitions, $S$ and $R$ are completely  specified by $T$ and $I$, with
for example
\begin{equation}
S=N - \frac{1}{2}(T + I+1) 
\end{equation}
The probability of an upward transition is $p_+=\alpha S/(\alpha S + N)$, whereas the probability of a downward transition is $p_-=1-p_+=N/(\alpha S + N)$.  These probabilities
are unequal and depend on $I$ and $T$, so that the walk is biased, with a "time-" and space-dependent drift.  (From here on, we will colloquially refer to $T$ as time, and trust this will not lead to confusion). The form of these probabilities simplify at threshold, $\alpha=1$, where $N-S$ and $I$ are both much smaller than $N$.  Then,
\begin{equation}
p_\pm = \frac{1}{2} \mp \frac{1}{8N}\left(T+I\right)
\end{equation}
where we have also assumed $T$ is large enough that we can ignore the 1.  Thus, the drift is, for large populations, very weak.  

This formulation immediately gives the well-known answer for an infinite population, where the bias term vanishes and we have a simple random walk starting at 1 with a trap at the
origin.  The distribution of first-passage times is 
\begin{equation}
P(T=2k+1)=2^{-2k-1}\left({{2k}\choose{k}}-{{2k}\choose{k+1}}\right)
\end{equation}
which for large $T$ becomes
\begin{equation}
P(T=2k+1)\approx\frac{1}{\sqrt{4\pi k^3}}
\end{equation}
Our task is to identify how the bias, resulting from the reduction of the susceptible pool with time, modifies this answer.

It is straightforward to generate the discrete-time master equation for our biased random walk. Since the bias is very weak, however, it is only effective at large times, and we are justified in passing to the Fokker-Planck equation for the distribution $P(I)$:
\begin{equation}
\frac{\partial}{\partial T} P(I,T)= \frac{1}{2}\frac{\partial^2}{\partial I^2}P + \frac{1}{4N}\frac{\partial}{\partial I}\left[\left(T+ I\right)P\right]
\end{equation}
One final simplification is to realize that the time-dependent drift is more effective than the spatially-dependent drift, and so the latter may be dropped.  The argument is straightforward: The typically "length" scale $I$ is proportional to $T^{1/2}$.  Thus,
the time-dependent drift is relevant when $T^{-1} \sim T^{1/2}/N$, or $T\sim N^{2/3}$.
The spatially-dependent drift become effective only when $T^{-1} \sim 1/N$ or $T \sim N$, much later than the time-dependent drift and so can be neglected.

\begin{figure}
\center{\includegraphics[width=0.45\textwidth]{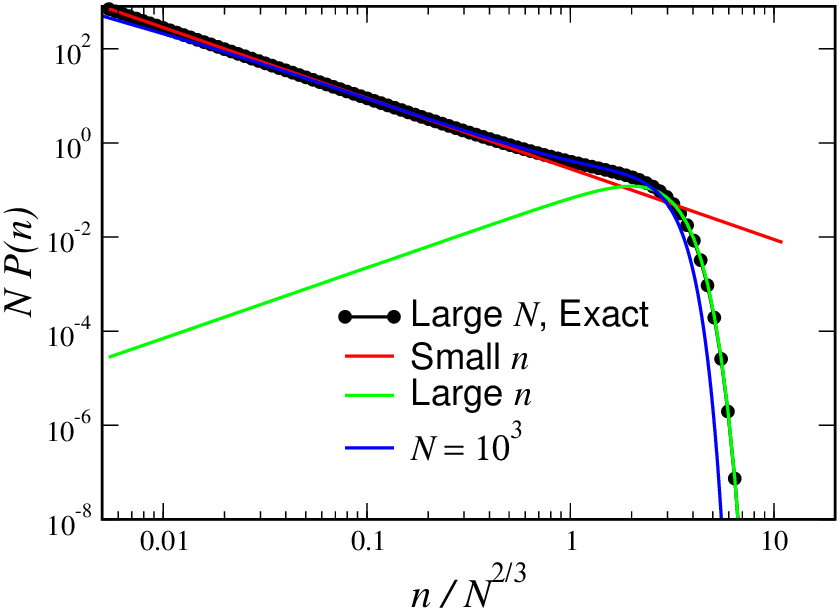}}
\caption{Scaled large-$N$ probability density $N P(n)$ for outbreaks of total size $n$, 
versus the scaled outbreak size $n/N^{2/3}$, from Eq. (\ref{exact}), together with
the small-$n$ (Eq. (\ref{smallT})) and large-$n$ (Eq. (\ref{largeT})) asymptotics. Also
displayed is the exact results for $N=1000$.}
\label{fig1}
\end{figure}

Thus the equation we need to solve is
\begin{equation}
\frac{\partial P}{\partial T}= \frac{1}{2}\frac{\partial^2 P}{\partial I^2} + \frac{T}{4N}\frac{\partial P}{\partial I}
\end{equation}
This equation is difficult to treat in its current form, since it is not separable, but becomes so if we define
\begin{equation}
P\equiv e^{-IT/4N-T^3/(96N^2)}\psi
\label{psi}
\end{equation}
so that
\begin{equation}
\frac{\partial \psi}{\partial T} = \frac{1}{2}\frac{\partial^2 \psi}{\partial I^2} + \frac{I}{4N}\psi .
\label{fpe}
\end{equation}
with the boundary conditions $\psi(0,T)=0$, $\psi(I,0)=\delta(I-1)$.  We can eliminate $N$
from the equation by the scaling $T\equiv 2a^2\tilde T$, $ I \equiv a \tilde I$, with  $a=(2N)^{1/3}$, resulting in (after dropping the tildes)
\begin{equation}
\frac{\partial \psi}{\partial T} = \frac{\partial^2 \psi}{\partial I^2} + I\psi .
\end{equation}
with $\psi(I,0)=\delta(I-1/a)/a$.
The operator on the right-hand side has an spectrum unbounded from above, so we need to regularize the problem by imposing an absorbing wall at some large $L$, which we will remove to infinity at the end.  Clearly, introducing such a wall in the original equation for $P$ has no significant effect, so it cannot materially affect our calculation in terms of $\psi$.  With this regularization, the right-hand operator has a well-defined discrete spectrum, with eigenvalues $E_n$ and normalized eigenfunctions $\phi_n$.  In terms of this, the flux of $\psi$ to the trap at the origin is given by
\begin{eqnarray}
{\cal{F}}_\psi&=&\frac{1}{2}\left.\frac{\partial \psi}{\partial I}\right|_{I=0}=\frac{1}{2a}\sum_n \phi_n'(0) \phi_n'(\frac{1}{a}) e^{E_n T}\nonumber\\
&\approx& \frac{1}{2a^2}\sum_n (\phi_n'(0))^2 e^{E_n T}
\end{eqnarray}
The eigenfunctions $\phi_n$ are given by
\begin{equation}
\phi_n(I)=A_n \text{Ai}(-x+E_n) + B_n \text{Bi}(-x+E_n)
\end{equation}
The condition $\phi_n(0)=0$ implies that
\begin{equation}
B_n = - A_n \text{Ai}(E_n)/\text{Bi}(E_n)
\end{equation}
and so, using the fact that the Wronskian of Ai and Bi is $1/\pi$,
\begin{equation}
\phi_n'(0)=-A_n/(\pi \text{Bi}(E_n))
\end{equation}
The normalization condition is
\begin{eqnarray}
1&=&\int_{0}^L \phi_n^2(I) dI  = \left[\phi_n'^2 + (x-E)\phi_n^2 \right]^L_0\nonumber\\
&=&\left[\left(\phi_n'(L)\right)^2 - \left(\phi_n'(0)\right)^2\right]
\approx \frac{(A_n^2 + B_n^2)L^{1/2}}{\pi}
\end{eqnarray}
where we have used the fact that $L$ is large to approximate Ai and Bi by their asymptotic expansions for large negative arguments.\cite{AbS}  The density of 
states is easily calculated from these expansions to be 
\begin{equation}
\frac{dn}{dE} \approx \frac{L^{1/2}}{\pi}
\end{equation}
We thus have, taking away the cutoff,
\begin{equation}
{\cal{F}}_\psi\approx \frac{1}{2\pi^2 a^2} \int_{-\infty}^{\infty} \frac{dE}{\text{Ai}^2(E)+\text{Bi}^2(E)}e^{ET}
\label{exact}
\end{equation}
Translating back to the original units and $P$, this gives our major result for the
probability density of extinction of the epidemic at transition $T$, with $S=N-T/2$ susceptible individuals left, so that $n=T/2$ individuals in all have been infected:
\begin{equation}
P(n)=\frac{e^{-n^3/(12N^2)}}{\pi^2 a^3}\int_{-\infty}^{\infty} \frac{dE}{\text{Ai}^2(E) + \text{Bi}^2(E)}e^{En/a^2}
\end{equation}
The first point to note is that this result satisfies the scaling behavior claimed by BN-K.
To understand this result in more depth, we compute its asymptotic behavior, first for
small $n$.  In this limit, the integral is dominated by the integrand at large negative $E$,
for which the integrand can be approximated by
\begin{equation}
\frac{1}{\text{Ai}^2(E)+\text{Bi}^2(E)}e^{ET/(2a^2)} \approx \pi (-E)^{1/2} e^{ET/(2a^2)}
\end{equation}
This gives the result
\begin{equation}
P(n) \approx \frac{1}{\sqrt{4\pi}n^{3/2}}
\label{smallT}
\end{equation}
as it should, since the drift is not relevant for small $n$.  For large $n$, the integral
is dominated by positive $E$'s of order $n^2$.  Then we  the integrand is dominated by Bi, and the integral becomes
\begin{eqnarray}
P(n)&\approx& \frac{1}{\pi^2 a^3} \int_{-\infty}^{\infty} \pi E^{1/2} e^{-\frac{4}{3} E^{3/2} + En/a^2}dE\nonumber\\ 
&\approx& \frac{1}{8\sqrt{\pi}N^2} n^{3/2} e^{-n^3/(16 N^2)}
\label{largeT}
\end{eqnarray}
Thus, $P(n)$ is sharply cut off for $n \gg O(N^{2/3})$ in accord with the numerical results of 
BN-K.
In Fig. \ref{fig1}, we graph $P(n)$, together with the asymptotic formulas for small 
and large $T$.  Also displayed is  the exact solution of the
full master equation for $N=10^3$.  We see that indeed the finite systems 
converge nicely to the scaling limit, with slower convergence at very small $n$, where
the discreteness of $n$ is relevant, and at the largest $n$ where our dropped spatially-dependent bias plays a detectable role.

\begin{figure}
\center{\includegraphics[width=0.45\textwidth]{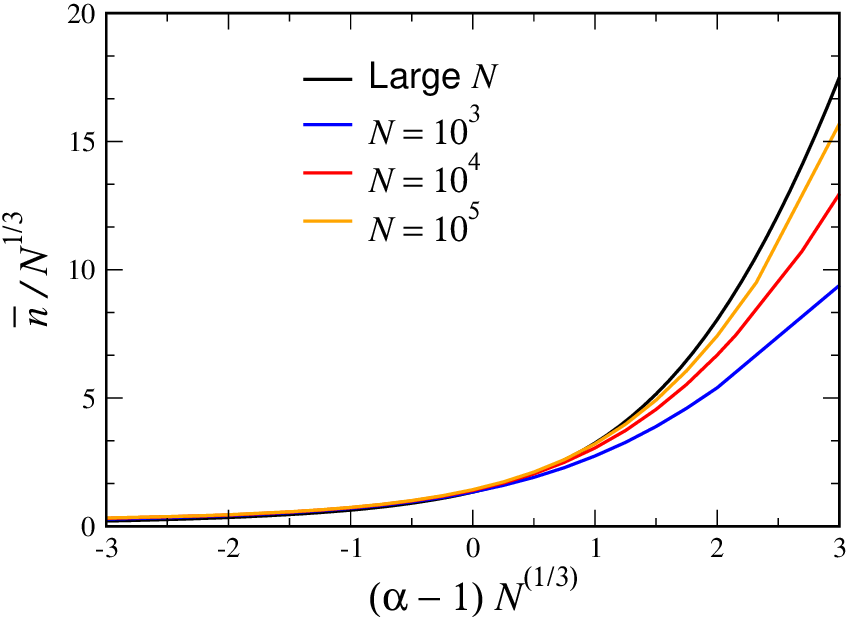}}
\caption{$\bar{n}/N^{1/3}$ as a function of $(\alpha-1)N^{1/3}$ for $N=10^3$, $10^4$,
and $10^5$, together with a numerical calculation of our large-$N$ analytic formula.}
\label{threshreg}
\end{figure}

\begin{figure}
\center{\includegraphics[width=0.45\textwidth]{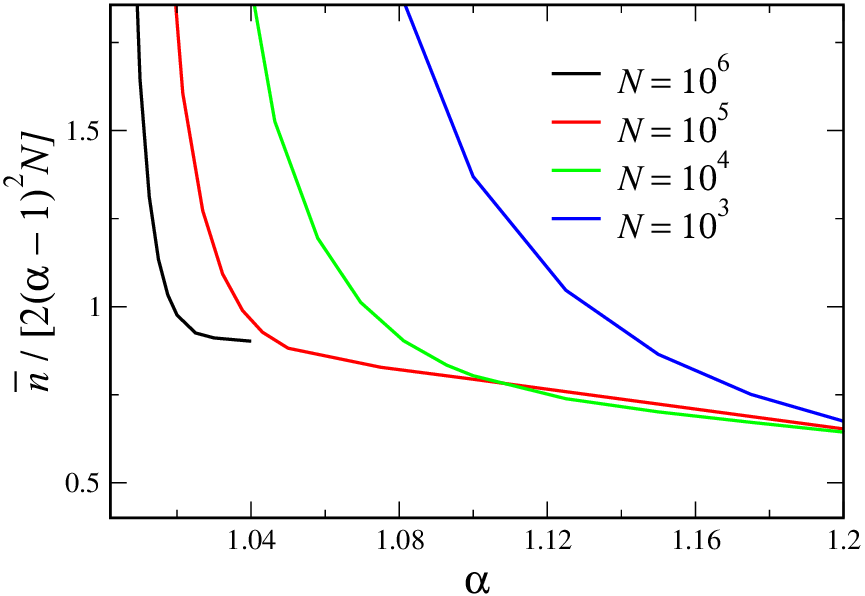}}
\caption{$\bar{n}/[2(1-\alpha)^2 N]$ as a function of $\alpha$, for $N=10^3$, $10^4$, 
$10^5$ and $10^6$.}
\label{above}
\end{figure}

\begin{figure}
\center{\includegraphics[width=0.45\textwidth]{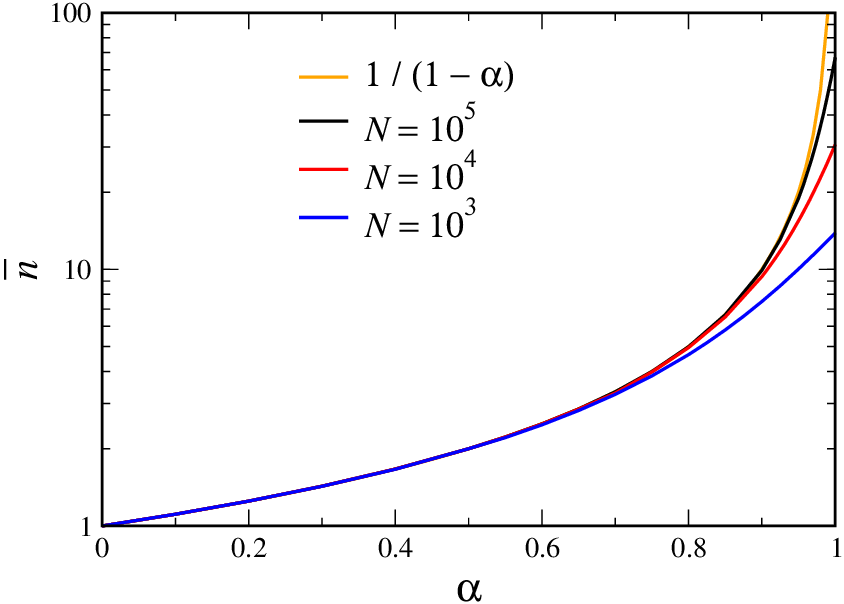}}
\caption{$\bar{n}$ as a function of $\alpha$, for $N=10^3$, $10^4$, and $10^5$, together with the infinite-$N$ prediction for the subcritical regime.}
\label{below}
\end{figure}

It is straightforward to extend our solution to the near threshold case, where
$\alpha = 1 + \delta$.  This introduces an additional constant bias to the problem.  Eq.
(\ref{fpe}) remains unchanged, where now $\psi$ is related to $P$ by
\begin{equation}
P\equiv e^{-\frac{I(T-2\delta N)}{4N}-\frac{(T-2\delta N)^3-(2\delta N)^3}{96N^2}}\psi ,
\end{equation}
so that in the probability distribution for outbreaks,  one gets an additional factor
 $\exp((n^2\delta N-n\delta^2N^2)/(3 N^2))$  The appropriate scale for $\delta$ is clearly
$N^{-1/3}$ as noted by BN-K.  In this context it is interesting to consider the average outbreak size as a function of $\delta$.  The scaling with $N$ of $P(n)$ implies that
the average outbreak $\bar{n}$ scales as $N^{1/3}$.  While $\bar{n}(\delta)$ is given by
a double integral, and must be computed numerically, the asymptotic behavior for large
positive and negative $\delta$'s is accessible.  For large positive $\delta$, the double integral has a sharp peak at $n=2\delta N$, $E=n^2/(4a^4)$, giving 
\begin{equation}
\bar{n}\approx 2\delta^2 N
\label{posasymp}
\end{equation}
This is exactly what is needed to match on to the supercritical regime.  
For $\alpha-1 \gg N^{-1/3}$, on average a finite fraction of the entire population is infected before the epidemic runs its course.  The probability that the epidemic survives to macroscopic proportions is $1-1/\alpha$,\cite{Getz} in which case the deterministic prediction of
$S_f$ is reliable.  Thus the average outbreak is of size
\begin{equation}
\bar{n}= \frac{\alpha-1}{\alpha} r N ,
\label{biga}
\end{equation}
where $r$ satisfies $r + e^{-\alpha r} = 1$, which for $\alpha$ near 1 yields Eq. (\ref{posasymp}).  The exact results from the master equation for the supercritical regime are in excellent agreement with Eq. (\ref{biga}), except near the threshold region $\alpha\approx 1$ (data not shown).
For large negative $\delta$, on the other hand, small $n$'s predominate, and
\begin{equation}
\bar{n}\approx \int_0^\infty  dn\, n\, e^{-n\delta^2/4} \frac{1}{2\sqrt{\pi}n^{3/2}} = 1/\delta
\end{equation}
This in turn matches on to the subcritical result, $\bar{n}=1/(1-\alpha)$ as $\alpha$
approaches one from below.  Thus, the near threshold hold regime interpolates smoothly between the sub- and super-threshold domains.  In the former, the probability distribution is sharply peaked at 0, whereas in the latter there are two peaks, one at zero and a second at the deterministic value of $n$. It is in the near-threshold regime that this second peak is born and splits off from the first.  In Fig. \ref{threshreg}, we plot $\bar{n}(\delta)$
obtained from a numerical integration of our formula, together with the results for $N=10^3$, $N=10^4$, and $N=10^5$.  We see
that the agreement is excellent for small $|\delta|$ where the results have converged.
Convergence is much slower, however, for large $\delta$.  To see the convergence to
the asymptotics better, we plot in Fig. \ref{above} $\bar{n}/(2\delta^2 N)$, and we see
that the data approach the expected value of 1 as $\delta$ approaches 0, only to veer
off as the ($N$-dependent) threshold regime is found.  Similarly, in Fig \ref{below}, we
show $\bar(n)$ together with the subcritical prediction, and see how the finite $N$
data follow the diverging curve as $\alpha$ approaches 1, only to level off in the threshold region.

In summary, we have found an analytic solution of the SIR model in the vicinity of the
infection threshold, we interpolates between the solutions of the sub- and super-critical cases.  This solution exhibits the scaling properties found by BN-K.  In addition, we have presented a solution to the problem of the first-passage of a random walker with a drift which increases linearly in time.

DAK acknowledges the support of the Israel Science Foundation.  The work of NMS is supported in part by the EU 6th framework CO3 pathfinder.

\end{document}